\documentclass{article}
\usepackage{natbib,epsfig,multirow}

\usepackage{amsthm,amsmath,amssymb,colonequals,multirow,graphicx,epstopdf}
\newcommand{\tr}{\,\mbox{tr}}

\newcommand{\matd}{\mathbf{D}}
\newcommand{\mata}{\mathbf{A}}
\newcommand{\matb}{\mathbf{B}}

\newcommand{\matf}{\mathbf{F}}

\newcommand{\matw}{\mathbf{W}}

\newcommand{\matp}{\mathbf{P}}
\newcommand{\matr}{\mathbf{R}}

\begin{document}

\title{Estimating Common Principal Components in High Dimensions}
\author{Ryan P.\ Browne\thanks{E-mail: rbrowne@uoguelph.ca. Tel: +1-519-824-4120, ext.\ 53034.} \ and~Paul~D.~McNicholas
}
\date{Department of Mathematics \& Statistics, University of Guelph.}

\maketitle

\begin{abstract}
We consider the problem of minimizing an objective function that depends on an orthonormal matrix. This situation is encountered when looking for common principal components, for example, and the Flury method is a popular approach. However, the Flury method is not effective for higher dimensional problems. We obtain several simple majorization-minizmation (MM) algorithms that provide solutions to this problem and are effective in higher dimensions. We then use simulated data to compare them with other approaches in terms of convergence and computational time.
\end{abstract}


\section{Introduction}

The minimization of the objective function 
	\begin{equation} \label{objective}
	 f(\matd) = \sum_{g=1}^G \tr\{\matw_g\matd\mata_g^{-1}\matd'\}
	\end{equation}
is required for a potpourri of statistical problems. To minimize an objective function $f(\matd)$ that depends on a orthonormal matrix $\matd$ (such as an eigenvector matrix), the search space is the orthogonal Stiefel manifold. This manifold is an embedded sub-manifold of $\mathbb{R}^{p \times p}$ equal to the set of all orthonormal matrices $\mathcal{M}= \left\{ \matd \in  \mathbb{R}^{p \times p} : \matd' \matd = \mathbf{I}_p\right\}$, where $\mathbf{I}_p$ denotes the $p$-dimensional identity matrix. The matrices $\matw_1,\ldots, \matw_G$ are positive-definite and are usually sample covariance matrices. The matrices $\mata_1,\ldots, \mata_G$ are diagonal matrices with positive elements. 

In \cite{flury88} a common principal components model for $G$ groups is found by minimizing \eqref{objective}. \cite{schott1998} and \cite{boik2003} use this objective function to find common principal components on correlation matrices. \cite{merbouha2004} use this objective function as a regularized technique in discriminant analysis with mixed discrete and continuous variables for generalized location models. \cite{yang2006} use this decomposition for multivariate time series data sets and use it in various multimedia, medical and Þnancial applications. \cite{celeux95} give an expectation-maximization (EM) algorithm \citep{dempster77} wherein each M-step the minimization \eqref{objective} is preformed. \cite{boik2002} note that the common principal components model has been employed in many fields such as the genetics, climatology, ontogeny, and other fields \citep{arnold1999, klingenberg1996, kulkarni2000, krzanowski1990, sengupta1998, oksanen1989}.

The Flury-Gautschi (FG) algorithm \citep{flury86} is the most popular algorithm to minimize \eqref{objective}.  \cite{lefkomtch1993} report that the FG ``is computationally expensive, especially for large and/or many matrices.''  \cite{boik2002} agree, pointing out that the Flury-Gautschi algorithm is slow in higher ($p > 5$) dimensions. \cite{browne2012} also show that the FG algorithm is slow in high dimensions. This limitation in application of the FG algorithm has had knock-on effects in methods that use it. For example, in a high dimensional mixture modelling application, \cite{bouveyron2007} avoid the common principal component models stating that they ``require a complex iterative estimation based on the FG algorithm \citep{flury86} and therefore they will be not considered here.'' To overcome a slow algorithm in high dimensions, \cite{browne2012} implemented an accelerated line search (ALS) for optimization on the orthogonal Stiefel manifold in a mixture modelling application and showed that this outperforms the FG method in high dimensions and reduces the number of degenerate solutions produced by the EM algorithm. In their ALS, \cite{browne2012} do not exploit the convexity of the objective function. In this paper, however, we exploit convexity to obtain several simple majorization-minizmation (MM) algorithms \citep[c.f.][]{hunter2000, hunter2003} following methodology from \cite{kiers2002}. We then compare all algorithms that minimize \eqref{objective} in terms of convergence and computational time.

\section{Minimization on the orthogonal Stiefel manifold}\label{sec:method}

\subsection{Flury Method} 

\cite{flury86} suggest an algorithm based on pairwise minimization of the matrix $\matd$.
That is, each pair of columns or eigenvectors of $\matd$ is updated while holding the others fixed. These updates are based on the eigendecompostion of $2\times2$ matrices summed across groups. Then we are required to loop through all pairs of columns of the matrix $\matd$ to complete a single iteration. This makes the Flury method ineffective in higher dimensions. See \cite{flury86} for details on the algorithm. 

\subsection{Accelerated Line Search} 

An accelerated line search algorithm (ALS) on a manifold consists of selecting a search direction in the tangent space and then moving this direction until a `reasonable' decrease in the objective function is found. \cite{browne2012} introduces an ALS algorithm to minimize the function in equation \eqref{objective}. An extensive review of optimization on matrix manifolds is given by \cite{AbsMahSep2008}. This methods requires tunning parameters and we use the values suggested by \cite{browne2012}.

\subsection{MM Algorithm 1} \label{MM 1}

We can exploit the convexity of the objective function  \eqref{objective} to obtain a MM algorithm similar that given in  \cite{kiers2002}. Three different MM algorithms are presented and each algorithm has a surrogate  function  of the form
\begin{equation*} 
	 f(\matd) = \sum_{g=1}^G \tr\{\matw_g\matd\mata_g^{-1}\matd'  \}
	 \le  C + \tr\left\{ \matf_t \matd \right\} 
\end{equation*}
where $C$ is a constant that does not depend on $\matd$, 
$\matf_t = \sum_{g=1}^G \left( \mata_g^{-1} \matd_t' \matw_g  -  \omega_k \mata_g^{-1} \matd_t' \right)$,
$\omega_g$ is largest eigenvalue of the matrix $ \matw_g$, and subscript $t$ denotes iteration number. The largest eigenvalue of a matrix can be determined using the power iteration method \citep{vonmises29}. If we obtain the singular value decomposition 
$\matf_t= \matp_t \matb_t \matr_t'$,
in which $\matp_t$ and $\matr_t$ are orthonormal, and $\matb_t$ is diagonal, containing the singular values of $\matf_t$ on the diagonal.
Then the update of the matrix $\matd$ becomes
$\matd_{t+1} = \matr_t \matp_t'$
Then we iteratively repeats this process until convergence.

\subsection{MM Algorithm 2} \label{MM 2}

In the second MM algorithm, 
$\matf_t = \sum_{g=1}^G \left( \matw_g   \matd_t \mata_g^{-1} -  \alpha_k  \matw_g \matd_t \right)$,
where $\alpha_g$ is largest eigenvalue of the matrix $ \mata_g^{-1}$. Because $\mata_g$ is diagonal and positive definite, the largest eigenvalue of  $\mata_g^{-1}$ is easily determined. The minimum of the surrogate is found using the same method as in Section~\ref{MM 1}.
 
\subsection{MM Algorithm 3} \label{EVE MM 3}
 
In the third MM algorithm,
$\matf_t = \sum_{g=1}^G \left( \matw_g   \matd_t \mata_g^{-1} -  \lambda_g  \matd_t \right)$,
where $\lambda_g$ is largest eigenvalue of the matrix $\mata_g^{-1} \otimes \matw_g$. Because the matrix $\mata_g^{-1} \otimes \matw_g$ is the Kronecker product of two matrices, we have $\lambda_g = \alpha_g \omega_g$. The minimum of the surrogate is found using the same method as in Section~\ref{MM 1}.
 
\subsection{MM Algorithm 4 for the EVE model} \label{EVE MM 4}
 
We iterate over MM algorithm 1 and MM algorithm 2.

\section{Simulation Study}\label{sec:sim}

We simulate various instances of the problem of minimizing \eqref{objective} to compare our approach to the Flury method and the accelerated line search used in \cite{browne2012}. We randomly generated $\matw_1,\ldots,\matw_G$ where each was produced from a $p+1$ observations from the $p$-dimensional standard normal distribution. In addition, we randomly generated the diagonal elements $\mata_1,\ldots,\mata_G$ from the half-normal distribution. Then we varied the number of dimensions~$p$. We used the identity matrix as a starting value for each algorithm and then we ran until convergence. For each simulation, we recorded the system time, the number of iterations, and value of the objective function at the converged solution. 

Table \ref{tab:p5and20} shows averages of the system times and the number of iterations from 100 simulations of the six algorithms. The table also gives the the relative difference between the minimum and the converged minimum (`\% Diff.') for each case. For a particular simulation, if $\{t_1, \ldots, t_6 \}$ are the values of the objective function from the converged solutions from the six algorithms and we let $t_{\text{min}} =\mbox{min}\{t_1, \ldots, t_6 \}$, then difference percentage for algorithm $k$ is $(t_k - t_{\text{min}})/t_{\text{min}}$, for $k=1,\ldots 6$. Note that if an algorithm has a large `\% Diff.' then we could use a stricter convergence criteria to improve the result. However, a stricter convergence criteria will also increase the number of iterations and thus the system time. To facilitate comparison, we used the same convergence criteria for each algorithm. 
\begin{table}[htb]
\caption{\label{tab:p5and20} The average system times (in seconds), iterations and difference between convergence value and minimum from the six algorithms.}
\begin{tabular*}{\textwidth}{@{\extracolsep{\fill}}c|crc|rrrl}
\hline
 &  \multicolumn{3}{c|}{p = 5, G = 5} & \multicolumn{3}{c}{ p = 20, G = 5} \\
Method & Time &  Iter. & \% Diff. & Time &  Iter. & \%  Diff.  \\
\hline
ALS  & 0.038&   34& 0.050& 0.292   &  83 & 8.549 \\
Flury & 0.050&   10& 0.011& 4.016    & 27 & 0.016  \\
MM 1& 0.055& 60& 0.007& 0.323   & 218 & 0.362 \\
MM 2& 0.027&  85& 0.010& 0.235 & 318  &1.045  \\
MM 3& 0.165&  179& 0.017& 0.872 & 580 & 2.526\\ 
MM 4& 0.035&  32& 0.001& 0.263  &  128 & 0.180 \\
\hline
\end{tabular*}
\end{table}
\begin{table}[htb]
\caption{\label{tab:p50and100} The average system times (in seconds), iterations and difference between convergence value and minimum from the six algorithms.}
\begin{tabular*}{\textwidth}{@{\extracolsep{\fill}}c|rrr|rrrl}
\hline
 &  \multicolumn{3}{c|}{p = 50, G = 5} & \multicolumn{3}{c}{ p = 100, G = 5} \\
Method & Time &  Iter. & \% Diff. & Time &  Iter. & \% Diff.     \\
\hline
ALS  & 2.85&   174& 0.347& 9.66   &  78 & 0.500 \\
Flury & 101.71&   33& 0.000&     &  &   \\
MM 1& 2.22& 303& 0.025& 10.51   & 290 & 0.010 \\
MM 2& 2.76&  565& 0.080& 22.12 & 836  &0.120  \\
MM 3& 8.62&  961& 0.187& 32.85 & 1521 & 0.355 \\ 
MM 4& 2.26&  214& 0.016 & 12.78  &  230 & 0.000 \\
\hline
\end{tabular*}
\end{table}

Table \ref{tab:p5and20} illustrates that the Flury method becomes computationally infeasible when the dimension increases from five to twenty. Also, it seems the ALS method and MM~3 algorithm tend to converge prematurely. Conversely, MM~4 seems to retain computational efficiency while maintaing the same convergence rate as the Flury algorithm. Table~\ref{tab:p50and100} tells the same story as Table~\ref{tab:p5and20} but with higher dimensions. Results for the Flury method are not reported in the right-hand column of Table~\ref{tab:p50and100} due to prohibitive computational time.  In the left-hand column of Table~\ref{tab:p50and100}, the Flury method converged to the smallest value in each simulation. The MM~4 gives similar results to the Flurry algorithm but is just as fast as the ALS algorithm. We note that the running parameters for ALS algorithm could be adjusted to optimize the performance of the ALS algorithm. 

\section{Discussion} \label{discussion}

We find that a MM algorithm is just as fast as the ALS algorithm introduced by \cite{browne2012} but has the same properties as the Flury method \citep{flury86}. In addition, the MM algorithm does not have any tunning parameters unlike the ALS algorithm. This will allow the implementation of techniques for higher dimensional problems for which the Flury method is too slow. Examples include the method of \cite{bouveyron2007} for clustering high dimensional data and parameter estimation for some of the mixture models considered by \cite{celeux95}.


\bibliographystyle{elsarticle-harv} 
\bibliography{essay} 

\end{document}